\documentclass{aa}

\usepackage{graphicx}
\usepackage{txfonts}
\usepackage[dvipsnames]{xcolor}
\usepackage{marvosym}
\usepackage{natbib}
\bibpunct{(}{)}{;}{a}{}{,}
\usepackage[none]{hyphenat} 
\usepackage{mathtools}
\usepackage{amsmath}
\usepackage{soul}
\usepackage{subcaption,booktabs}
\usepackage{multirow}
\usepackage{comment}
\usepackage{lscape}
\usepackage[thinc]{esdiff}

\definecolor{darkblue}{RGB}{12,94,176}
\usepackage{hyperref}
\hypersetup{hidelinks}
\hypersetup{
    pdfnewwindow=true,   
    colorlinks=true,      
    linkcolor=darkblue,   
    citecolor=darkblue,    
    filecolor=darkblue,   
    urlcolor=blue         
}

\usepackage{siunitx}

\DeclareSIUnit\parsec{pc}
\DeclareSIUnit\cMpc{cMpc}
\DeclareSIUnit\year{yr}
\DeclareSIUnit\Zsun{Z_{\odot}}
\DeclareSIUnit\Msun{M_{\odot}}
\DeclareSIUnit\Rsun{R_{\odot}}
\DeclareSIUnit\Lsun{L_{\odot}}
\DeclareSIUnit\erg{erg}
\DeclareSIUnit\eV{eV}

\newcommand{\orcidicon}[1]{\href{https://orcid.org/#1}{\includegraphics[width=11pt]{plots/ORCIDiD_icon128x128.png}}}

\begin{document}

   \title{Investigating the formation channel of GW231123: Population III stars or hierarchical mergers?}

   \author{
   Federico Angeloni\inst{1,2,3,4} \thanks{\email{federico.angeloni@students.uniroma2.eu}}
   \and
   Konstantinos Kritos\inst{5}
   \and
   Raffaella Schneider\inst{1,3,4}
   \and
   Emanuele Berti\inst{5}
   \and 
   Luca Graziani\inst{1,3,4}
   \and
   Stefano Torniamenti\inst{6, 7}
   \and
   Michela Mapelli\inst{7, 8}
   \and
   Ataru Tanikawa\inst{9}
   }

   \institute{
   Dipartimento di Fisica, Sapienza, Università di Roma, Piazzale Aldo Moro 5, 00185, Roma, Italy\and
   Dipartimento di Fisica, Tor Vergata, Università di Roma, via della Ricerca Scientifica, 00133, Roma, Italy\and
   INAF/Osservatorio Astronomico di Roma, Via di Frascati 33, 00078 Monte Porzio Catone, Italy\and
   INFN, Sezione di Roma I, Piazzale Aldo Moro 2, 00185 Roma, Italy\and
   William H. Miller III Department of Physics and Astronomy, Johns Hopkins University, 3400 N. Charles Street, Baltimore, Maryland, 21218, USA\and
   Max-Planck-Institut für Astronomie, Königstuhl 17, 69117, Heidelberg, Germany\and
   Universität Heidelberg, Zentrum für Astronomie (ZAH), Institut für Theoretische Astrophysik, Albert Ueberle Str. 2, 69120, Heidelberg, Germany\and
   Physics and Astronomy Department Galileo Galilei, University of Padova, Vicolo dell’Osservatorio 3, 35122 Padova, Italy\and
   Center for Information Science, Fukui Prefectural University, 4-1-1 Matsuoka Kenjojima, Eiheiji-cho, Fukui 910-1195, Japan
   }

   \date{Received ...; accepted ...}
 
  \abstract
   {The gravitational wave event GW231123, with component black hole masses lying within or above the pair-instability mass gap, poses a significant challenge to current stellar evolution models. In this work, we describe how we investigated its origin by coupling the galaxy formation model \texttt{GAMESH} with the cluster population synthesis code \texttt{RAPSTER} and using two distinct binary population synthesis codes, \texttt{SEVN} and \texttt{BSEEMP}. This framework allowed us, for the first time, to reconstruct the life cycle of GW231123-like candidates within the same cosmological simulation, enabling a self-consistent comparison between different formation channels.
   Although both population synthesis codes can in principle produce black holes compatible with GW231123, we find that isolated binary evolution fails to reproduce the inferred merger redshift. In \texttt{SEVN}, massive black hole binaries form with semi-major axes $>10^3\, \rm R_\odot$, preventing coalescences within a Hubble time. In \texttt{BSEEMP}, candidates arise only at extremely low metallicities ($ \rm Z \approx 10^{-10}$), which contribute negligibly to the star formation rate density in our overdense simulated volume.
   Our results therefore strongly support a dynamical hierarchical origin. The observed black hole masses are naturally reproduced through successive mergers in dense globular clusters. The high dimensionless spins reported by the LIGO-Virgo-KAGRA Collaboration are consistent with this hierarchical population. We found a local merger rate density of $0.78 \, \rm Gpc^{-3}\,yr^{-1}$, with a  peak at \textit{z} $= 4-6$, tracing the maximum formation rate of globular clusters in metal-poor environments ($\rm Z\approx0.006$). Overall, GW231123 may represent a benchmark event for a robust population of hierarchical black holes formed in the early Universe.}

    \maketitle

\section{Introduction}
The recent detection of GW231123 \citep{LIGO+2025arXiv} has introduced a new class of gravitational-wave sources. This high-confidence event \citep{Bini+2026arXiv260109678B} involves the merger of two black holes (BHs) with source-frame masses of $\rm M_1=137^{+22}_{-17}$ $\rm M_\odot$ and $\rm M_2=103^{+20}_{-52}$ $\rm M_\odot$ at a redshift of \textit{z} = $0.39^{+0.27}_{-0.24}$ \citep{Abac+2025ApJ}. Its large total mass -- which places GW231123 within the pair-instability (PI) mass gap -- and the exceptionally high spins ($\chi_{1} \approx 0.9$ and $\chi_{2} \approx 0.8$, see \citealt{Abac+2025ApJ}) challenge standard astrophysical formation scenarios and waveform modeling. \\
The PI mass gap, typically predicted to span $\sim 60-130$ $\rm M_\odot$ \citep{Spera+2017MNRAS, Farmer+2019ApJ, Woosley+2021ApJ, Hendriks+2023MNRAS}, arises from PI supernovae that disrupt massive stellar cores. Although its exact boundaries depend on nuclear reaction rates and stellar rotation \citep{Farmer+2020ApJ, Mapelli+2020ApJ, Hirschi+2025MNRAS}, forming
such massive BHs remains difficult within conventional isolated evolution. 
Several formation channels have been proposed, including dynamical formation in AGN disks \citep{Delfavero+2025arXiv250813412D}, primordial BHs \citep{DeLuca+2025arXiv250809965D, Yuan+2025PhRvD} and cosmic strings \citep{Cuceu+2026PhRvD}. Among astrophysical scenarios, two leading channels can naturally account for the observed masses. \\
In dense stellar systems, such as globular clusters (GCs;  \citealt{Angeloni+2025arXiv,Tiwari+2026arXiv260205645T, Banerjee+2026arXiv260209694B, Martinez+2026}) and nuclear star clusters \citep{Newton+2026arXiv260204176N}, BHs can undergo successive mergers, progressively populating the PI mass gap \citep{Miller+2002MNRAS, Rodriguez+2019PhRvD, Paiella+2025ApJ}. This hierarchical process also explains the high spin values, as merger remnants inherit a dimensionless spin of $\sim$ 0.7 \citep{Berti+2008ApJ, Gerosa+2021NatAs}. However, it requires sufficiently massive clusters to retain remnants against gravitational wave recoil kicks \citep{Campanelli+2007ApJ, ArcaSedda+2023MNRAS}.\\
Alternatively, isolated Population III  (Pop~III) binaries -- first-generation stars in the early Universe -- may collapse directly into massive BHs due to their low metallicity and weak winds \citep{Croon+2025arXiv250810088C, Gottlieb+2025ApJ}. 
Yet reproducing both the large masses and high spins remains difficult, often requiring additional assumptions on stellar physics or accretion \citep{Liu+2025ApJ, Popa+2025ApJ, Stegmann+2025ApJ, Tanikawa+2025arXiv, Bartos+2026ApJ, Roupas+2026arXiv}. \\
In this paper we present the first self-consistent comparison of these two potential GW231123 formation channels within a common cosmological galaxy-formation framework, where binaries are populated according to the star formation rate (SFR) and metallicity of individual halos.
Section~\ref{Sec:Methods} outlines the model, while Sects.~\ref{Sec:Results} and \ref{sec:discussion} present the results and discussion.
\section{Methods}
\label{Sec:Methods}

\subsection{Galaxy formation model \texttt{GAMESH}}
\label{Sec:GAMESH}

Our cosmological framework relies on the \texttt{GAMESH} pipeline \citep{Graziani+2015MNRAS, Graziani+2017MNRAS}, which couples \texttt{N}-body dark matter dynamics (\texttt{GCD+} code; see \citealt{Kawata+2013MNRAS}) with semi-analytic prescriptions of star formation. This includes radiative and mechanical feedback and reionization-driven suppression in mini halos ($\rm T_{\rm vir} <  10^4\,\rm K$) at \textit{z} $\leq6$. We employed a multi-resolution zoom-in technique focusing on a  $(4\, \rm cMpc)^3$ comoving volume centered on a Milky Way (MW)-like halo of mass $\rm M_{\rm MW}$ = 1.7$\times$ $10^{12}$ $\rm M_\odot$ at \textit{z} = 0, representative of a Local Group (LG)-like environment. The model was calibrated against the MW's observed properties \citep{Graziani+2017MNRAS, Ginolfi+2018MNRAS}, and it provides a star formation history consistent with that inferred for the LG (see Fig.\ref{fig:SFvsGC_RD}).

\subsection{Population synthesis codes and coupling with \texttt{GAMESH}}
\label{BPS&CPS}

To model binary star evolution, we coupled \texttt{GAMESH} with binary population synthesis (BPS) codes. For each galaxy, we randomly drew binaries from the BPS databases until their cumulative mass matches the newly formed stellar mass predicted by the simulation, assuming progenitors inherit the metallicity of their host galaxy \citep{Schneider+2017MNRAS, Graziani+2020MNRAS}. \\
We employed two BPS codes to assess the impact of different stellar evolution prescriptions: \texttt{BSEEMP} \citep{Tanikawa+2021ApJ} and \texttt{SEVN} \citep{Iorio+2023MNRAS}. We assumed a binary fraction of $\rm f_b=1$ and adopted a Kroupa initial mass function (\citealt{Kroupa+2001MNRAS}) in the mass range $0.1 \,\rm M_\odot - 200 \, \rm M_\odot$, except for Pop~III stars, for which we assumed a log-flat initial mass function in the range $5 \, \rm M_\odot - 300 \, \rm M_\odot$ \citep{Hirano+2015MNRAS}. In Appendix \ref{Sec:PopIII_discussion} we provide more details on the adopted initial conditions to generate the databases. \\
Following \citet{Angeloni+2025arXiv}, GC formation was modeled by selecting galaxies with gas surface densities ${\Sigma_{\rm g} >  10^3\, \rm M_\odot\, \rm pc^{-2}}$ \citep[see][for reviews]{Adamo+2020SSRv, Kruijssen+2026enap}. For these systems, the cluster formation efficiency ($\Gamma$) was computed as in \citet{Kruijssen+2012MNRAS} to estimate the fraction of stellar mass bound in clusters. The GC masses were drawn from a $-2$ power-law initial mass function \citep{Li+2017ApJ} over $10^5-10^7\, \rm M_\odot$, with initial metallicities inherited from their host galaxies. Initial half-mass radii were sampled uniformly from a range between $0.1-1\,\rm pc$.
Their subsequent dynamical evolution and disruption -- driven by mass loss, tidal stripping, and galaxy mergers -- were modeled self-consistently using semi-analytic and \texttt{RAPSTER} prescriptions \citep{Angeloni+2025arXiv}.

\section{Results}
\label{Sec:Results}

\subsection{Isolated binary evolution channel for GW231123}

\begin{figure*}[t]
     \centering
     \begin{subfigure}[b]{0.45\textwidth}
         \centering
         \includegraphics[width=\textwidth]{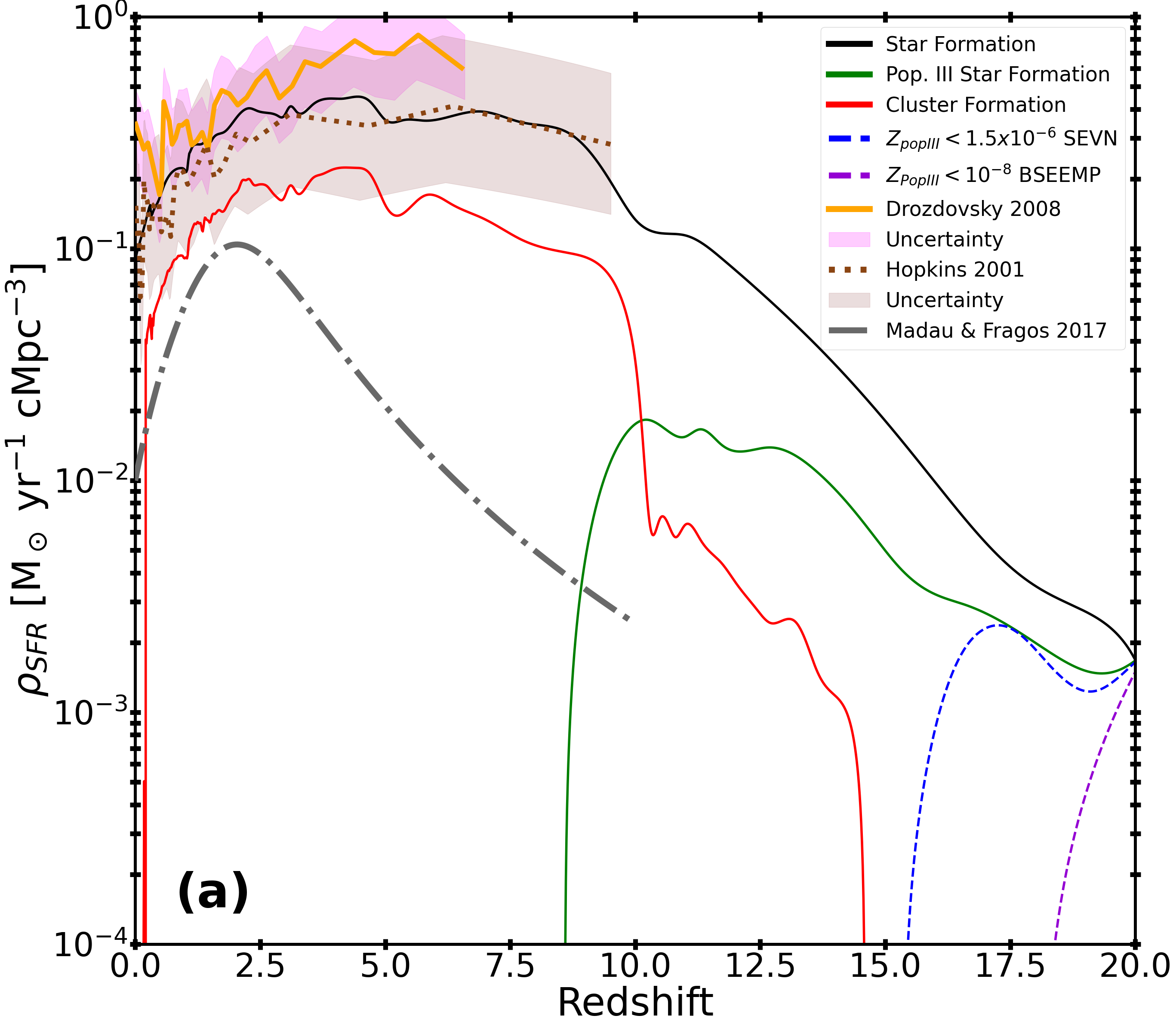}
         \phantomcaption
         \label{fig:SFvsGC_RD}
     \end{subfigure}
     \hfill
     \begin{subfigure}[b]{0.44\textwidth}
         \centering
         \includegraphics[width=\textwidth]{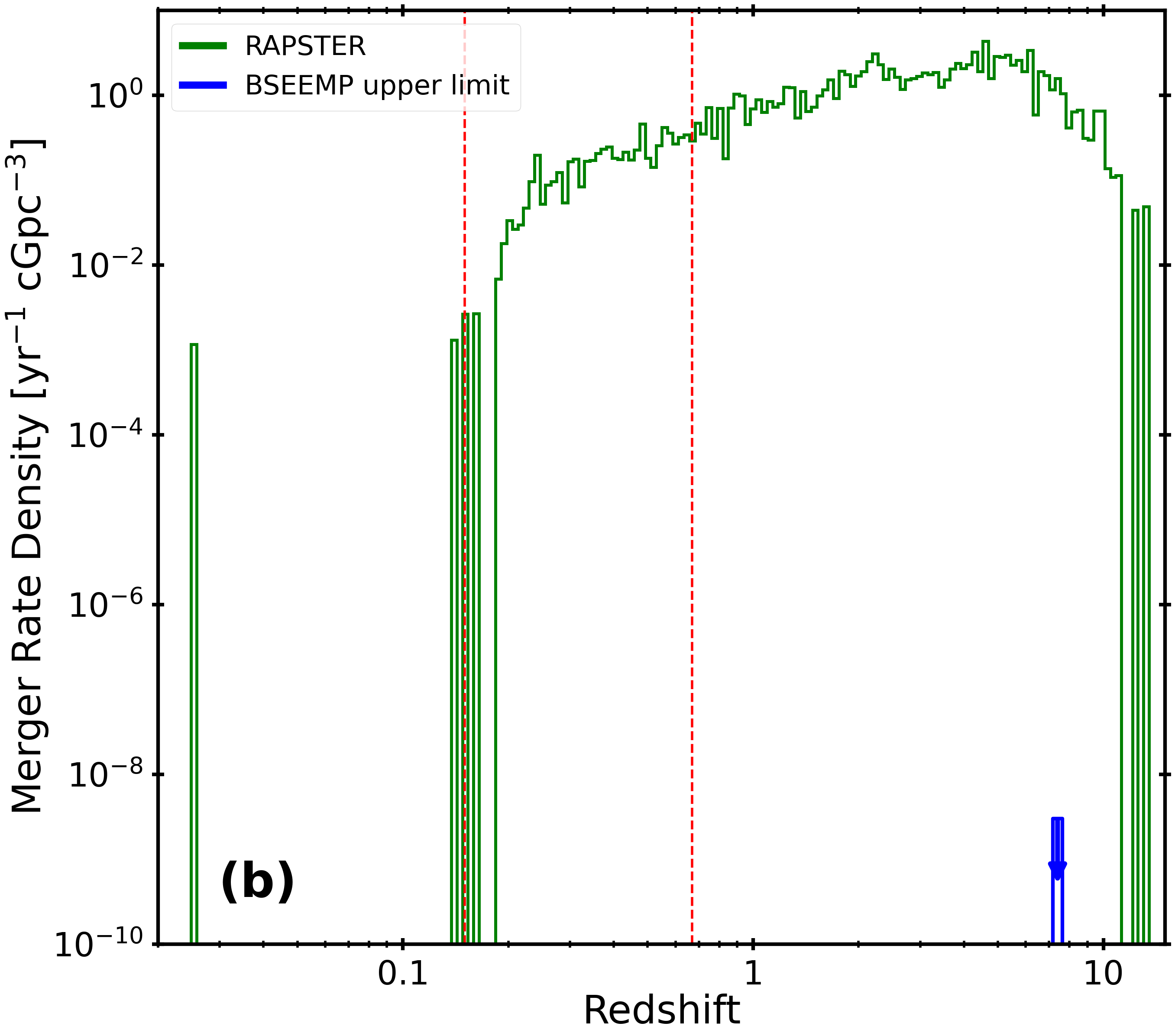}
         \phantomcaption
    \label{fig:MergerRateDensity}
     \end{subfigure}
     
        \caption{(a) Cosmological evolution of the total SFR density (black line), the Pop~III SFR density (green line), and the GC formation rate density (red line) predicted by \texttt{GAMESH}. The dashed lines identify the densities of Pop~III star-forming regions where BBHs with masses compatible with GW231123 (GW231123-mass BBHs) could form according to the two BPS codes. Models are compared with \citet[brown]{Hopkins+2001ApJ} and \citet[orange]{Drozdovsky+2008ASSP} data, including their uncertainty ranges, and with \citet[gray]{Madau+2017ApJ} cosmic SFR density evolution. (b) Merger rate density evolution of GW231123-mass BBHs within the simulated volume as predicted by RAPSTER (green) and the upper limit on the Pop III contribution from \texttt{BSEEMP} (blue). The two vertical dashed red lines indicate the 90\% credible interval for the merger redshift of GW231123 reported by the LVK Collaboration.}
        \label{}
\end{figure*}

We first assessed the contribution of isolated binary evolution by coupling BPS prescriptions with \texttt{GAMESH}. Although both codes can produce Pop~III binary BHs (BBHs) with masses compatible with GW231123, they fail to reproduce observable mergers at the inferred redshift within our cosmological volume. In \texttt{SEVN}, the most massive stellar progenitors ($\rm M_\star\sim240-300\,\rm M_\odot$) at $\rm Z < 10^{-6}$ form BBHs with semimajor axes exceeding $10^3\, \rm R_\odot$, preventing coalescence within a Hubble time. In \texttt{BSEEMP}, mergers arise only at the lowest metallicity  ($\rm Z \approx 10^{-10}$; see the violet line in Fig. \ref{fig:SFvsGC_RD}), but their formation efficiency is too low ($\ll 1\%$\footnote{The formation efficiency is defined as the ratio between the number of GW231123-like candidates ($\approx 40$) and the total number of binary systems simulated in the BPS catalog for a given metallicity, i.e., $2\times 10^7$.}), and their mean merger delay timescale is too short ($\simeq 0.5$ Gyr) to yield detectable events (see the upper limit in Fig. \ref{fig:MergerRateDensity} and Appendices \ref{Sec:PopIII_discussion} and \ref{app:BSEEMP_merger_rate} for further details). 

\subsection{Cluster formation and binary black hole merger rate}
\label{Sec:BBH rates}

Figure \ref{fig:SFvsGC_RD} shows the redshift evolution of the SFR and GC formation rate densities in the simulated volume. While the GC formation rate broadly follows the global star formation history, it declines more steeply at \textit{z} $> 10$ and \textit{z} $< 1$, indicating that conditions favorable for GC formation are less common at these epochs.\\
Figure \ref{fig:MergerRateDensity} shows the merger rate densities of GW231123-like candidates within our LG-like volume. The vertical dashed red lines indicate the inferred merger-redshift range of GW231123. Candidates were selected from the \texttt{RAPSTER} catalogs by requiring BH masses consistent with the $90\%$ credible intervals reported by the LIGO-Virgo-KAGRA (LVK) Collaboration.\footnote{Hereafter we refer to these systems as GW231123-mass BBHs.}  We found substantial overlap between our predicted mergers and the LVK redshift constraints. Moreover, the merger chains exhibit remarkable short delay times ($\sim 100\,\rm Myr$), resulting in a close correspondence  between the birth and merger rates. In a hierarchical scenario, this efficiency highlights the key role of dynamical interactions in dense clusters in driving BBHs to coalescence.
Reflecting on the overdense nature of our LG-like volume, we predict a local merger rate density of $0.78 \, \rm Gpc^{-3}\,yr^{-1}$ for GW231123-like candidates, which is approximately an order of magnitude higher than the value inferred by the LVK Collaboration ($0.08^{+0.19}_{-0.07}\, \rm Gpc^{-3}\, \rm yr^{-1}$, \citealt{Abac+2025ApJ}). This discrepancy is naturally explained by the enhanced star formation activity in our simulated region, whose $z \simeq 0$ SFR density exceeds the cosmic average by a factor of roughly eight (cosmic value: $\simeq  0.015 \, \rm M_\odot \, yr^{-1} \, Mpc^{-3}$, \citealt{Madau+2014ARAA}). \\
Owing to the short delay times of the hierarchical merger chains producing BBHs with masses comparable to GW231123 (hereafter GW231123-mass BBHs), the merger rate evolution shown in Fig. \ref{fig:MergerRateDensity} closely traces the formation history of these BBHs. They predominantly originate at $z\!\gtrsim\!2$, with a peak at $z\!\approx\!4-6$. This suggests that hierarchical assembly toward GW231123-mass BBHs is favored in the dense primordial environments of early-forming halos rather than in low-redshift GCs. Further details on the birth sites of GW231123-mass BBHs are provided in Appendix \ref{app:birth_places}.

\subsection{Black hole generation and metallicity distribution}

\begin{figure*}[t]
     \centering
     \begin{subfigure}[b]{0.45\textwidth}
         \centering
         \includegraphics[width=\textwidth]{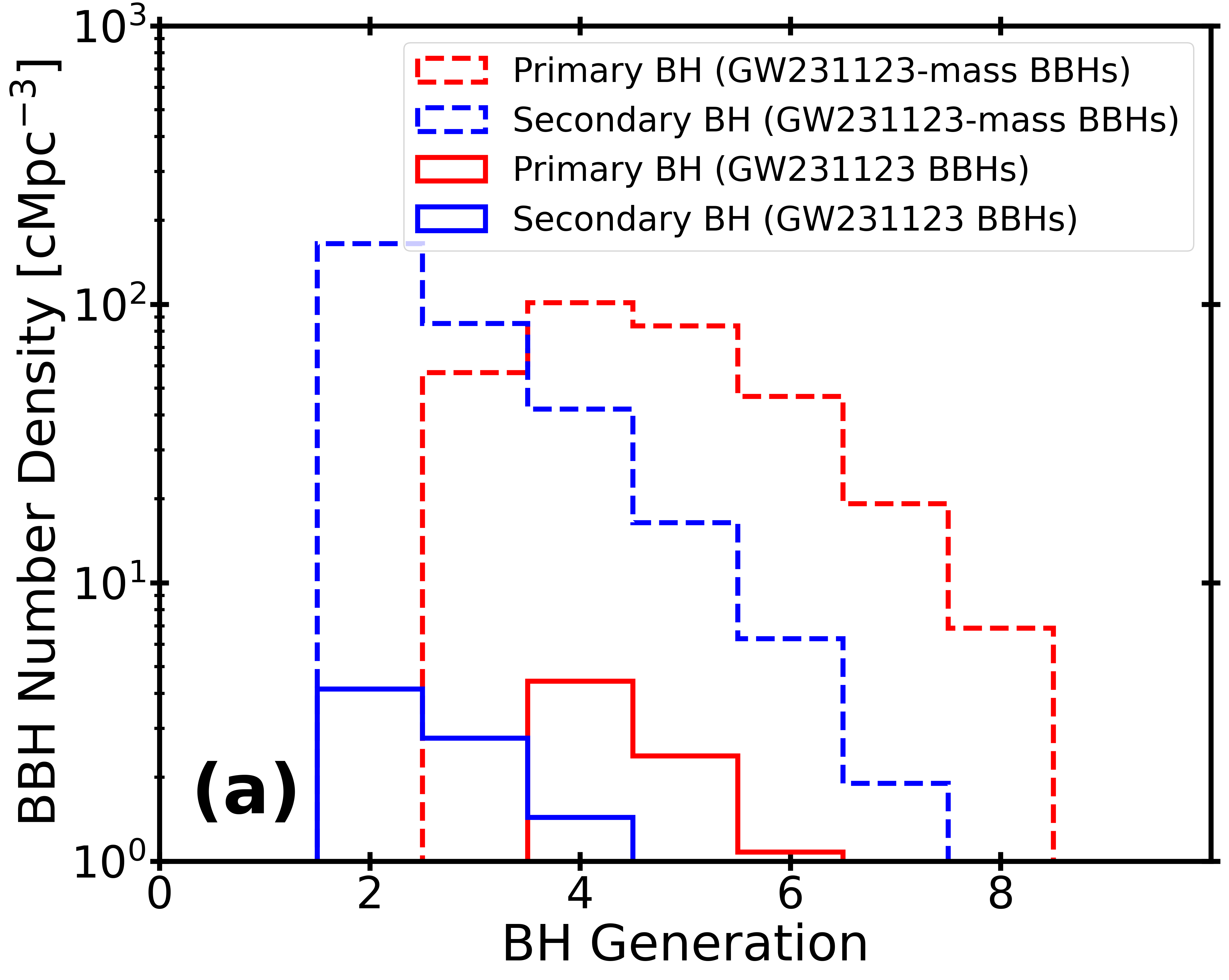}
         \phantomcaption
         \label{fig:BH_Gen}
     \end{subfigure}
     \hfill
     \begin{subfigure}[b]{0.47\textwidth}
         \centering
         \includegraphics[width=\textwidth]{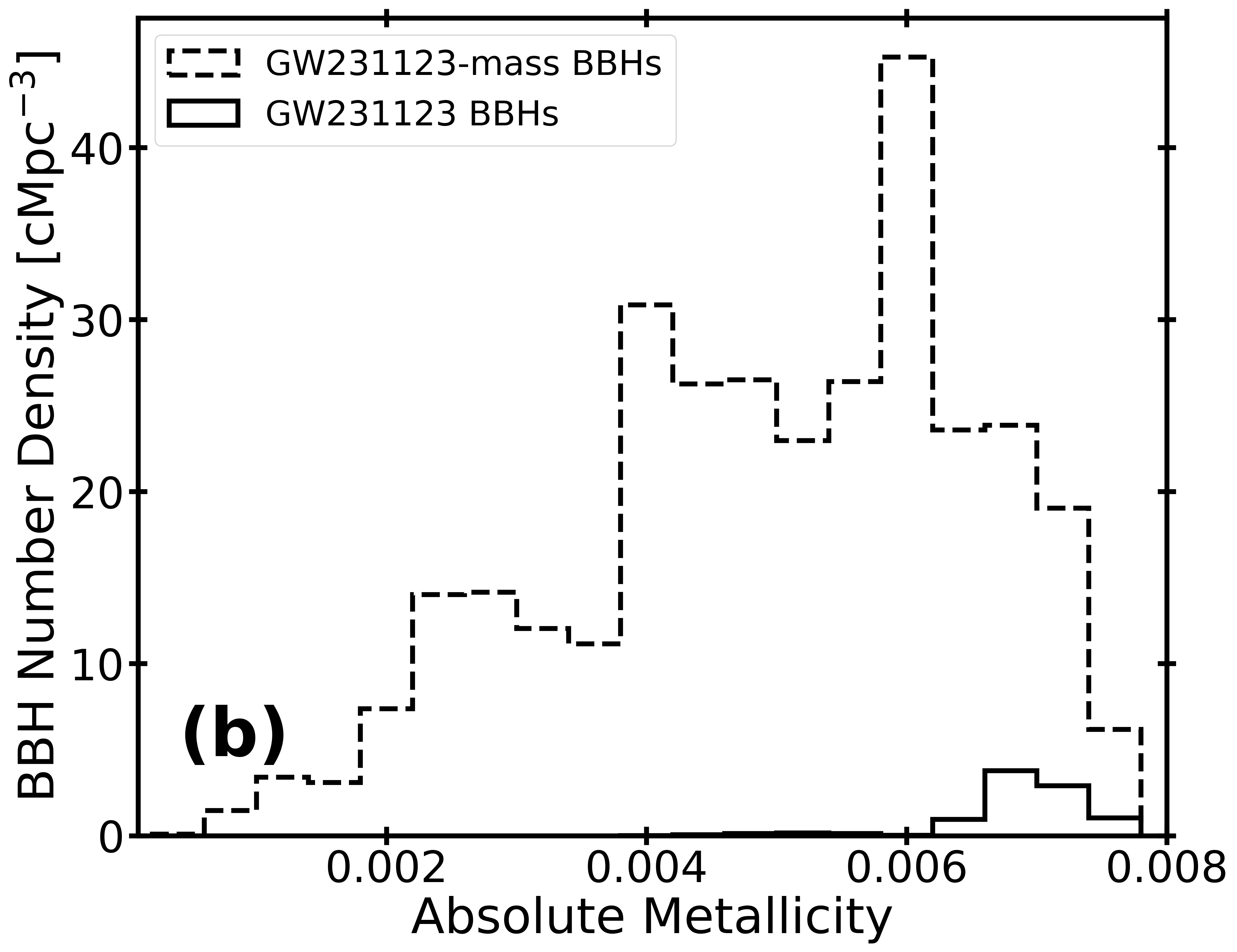}
         \phantomcaption
         \label{fig:Metallicity}
     \end{subfigure}
     
        \caption{(a) Binary black hole number density as a function of hierarchical merger generation for the primary (red) and secondary (blue) components of GW231123 candidates in our simulated volume. Dashed lines show systems selected only by the LVK mass priors (GW231123-mass BBHs), while solid lines indicate the fully consistent subset satisfying the LVK mass, merger-redshift, and spin constraints (GW231123 BBHs). (b) Metallicity distribution of GW231123-mass BBHs (dashed outline) and GW231123 BBHs (solid line) formation environments.}
        \label{fig:BH_Gen&Metallicity}
\end{figure*}

Figure \ref{fig:BH_Gen&Metallicity} illustrates the physical properties of the BBH populations identified as GW231123 candidates within our \texttt{RAPSTER}-\texttt{GAMESH} framework. Panel (a) shows the number density of primary and secondary components as a function of their hierarchical generation for GW231123-mass BBHs; the components were selected to match the LVK $\rm M_1$ and $\rm M_2$ posteriors. The primary BHs extend to higher generations (fourth to sixth), whereas secondaries are typically confined to the second and third generation. Imposing additional merger-redshift and spin constraints defines the consistent subset of GW231123 BBHs (solid lines), reducing the number density by about one order of magnitude while preserving the overall hierarchical trend (see also Appendix \ref{app:spin_analysis} for the spin distributions).\\
Panel (b) shows the formation-site metallicity distribution of GW231123 candidates. The distribution peaks at subsolar metallicities, $Z\!\simeq\!0.004\!-\!0.007$, indicating that massive hierarchical mergers preferentially form in metal-poor environments. However, a clear selection effect emerges when comparing the two subsets: GW231123-mass BBHs (dashed) span a broad range, $Z\!\approx\!0.001\!-\!0.008$, whereas the fully consistent GW231123 BBHs (solid) are confined to $Z\!>\!0.006$.
This shift toward higher metallicities arises because systems matching the observed merger redshift, combined with short delay times, must form in environments that have already undergone a more significant chemical enrichment at that epoch.

\section{Discussion}
\label{sec:discussion}

Our results strongly favor a hierarchical origin in which repeated mergers in GCs naturally reproduce the properties of GW231123-like binaries within a cosmologically consistent framework. \citet{Paiella+2025ApJ} identify nuclear clusters and GCs as primary sites for BBH build-up, though they favor very low metallicities ($Z\!\leq\!0.002$) and mainly low-order mergers. In contrast, we find that higher generations (fourth to sixth for the primary) are efficiently reached in massive GCs ($>\!10^5 \,\rm  M_\odot$) and that imposing the merger-redshift constraint selects systems at $Z\!>\!0.006$. This shift follows naturally from the short delay times ($\sim$100 Myr), which require formation in already enriched environments at the relevant epoch. \\
Our inferred host-cluster masses and retention requirements are consistent with recent dynamical studies. \citet{Mai+2025arXiv251021916M} show that only massive GCs ($\rm M_{\rm cl}\!>\!10^6 \, M_\odot$) can retain high-generation remnants against gravitational-wave recoil, while \citet{Tiwari+2026arXiv260205645T} infer preferred host masses of $10^{5.7}$–$10^{7.7} \,\rm  M_\odot$, which is in good agreement with our range (see Appendix \ref{app:birth_places}). Although \citet{Liu+2025ApJ} demonstrate that GW231123 could form in Pop~III star clusters ($\sim10^4\,\rm M_\odot$), our results, together with \citet{Banerjee+2026arXiv260209694B}, show that hierarchical mergers in more massive GCs can naturally populate the PI mass gap even at nonzero metallicity.

\section{Conclusions}

The detection of GW231123 challenges current interpretations of the PI mass gap and the formation of very massive BHs. By coupling the \texttt{GAMESH} cosmological framework with binary and cluster evolution models, we performed a self-consistent comparison between isolated Pop~III binary evolution and hierarchical assembly in dense stellar systems. Our results strongly favor a hierarchical origin in which repeated mergers in globular clusters naturally reproduce the properties of GW231123-like binaries. We find that BBHs with masses comparable to GW231123 predominantly merge at $z \approx 4-6$, implying that GW231123 represents the low-redshift tail of a broader population of hierarchical BHs formed in the early Universe. Future gravitational wave observatories will be crucial to revealing this hidden population and further constraining its astrophysical origin.

\begin{acknowledgements}
We are grateful to J. Bland-Hawthorn and A. Wetzel for their insightful comments.
FA, RS, and LG acknowledge support from the MUR projects FIS-2024-01621 DAWN and PRIN 2022CB3PJ3-FLAGS, from EU-Recovery Fund PNRR, and from the INFN TEONGRAV initiative.
KK and EB are supported by NSF Grants No.~AST-2307146, No.~PHY-2513337, No.~PHY-090003, and No.~PHY-20043, by NASA Grant No.~21-ATP21-0010, by John Templeton Foundation Grant No.~62840, by the Simons Foundation [MPS-SIP-00001698, E.B.], by the Simons Foundation International [SFI-MPS-BH-00012593-02], and by Italian Ministry of Foreign Affairs and International Cooperation Grant No.~PGR01167. Part of this work was carried out at the Advanced Research Computing at Hopkins (ARCH) core facility (\url{https://www.arch.jhu.edu/}), which is supported by the NSF Grant No. OAC-1920103. 
\end{acknowledgements}

\bibliographystyle{aa}
\bibliography{References}

\begin{appendix} 

\section{Initial conditions adopted to model Pop~III binary stars}
\label{Sec:PopIII_discussion}

In this section we describe the assumptions made to simulate populations of Pop~III binary stars with \texttt{SEVN} and \texttt{BSEEMP}. In particular, we outline the initial mass functions, stellar metallicity, and orbital parameter distributions adopted in our models.\\
In \texttt{GAMESH}, Pop III stars can form in star-forming regions with metallicity below a critical value of $Z_{\rm cr} = 2\times10^{-6}$. The two BPS codes explore different metallicity values within the Pop~III regime. \texttt{SEVN} considers $Z={2\times10^{-6},\,10^{-6},\, 10^{-11}}$, whereas \texttt{BSEEMP} adopts $Z={2\times10^{-6},\,2\times10^{-8},\,2\times10^{-10}}$. Each metallicity bin contains $2\times10^7$ binary systems to ensure optimal Monte Carlo sampling of initial conditions and a robust statistical representation within cosmological volumes.\\
We sample primary stellar components ($\rm M_{1\star}$) within the range $[5-300]\,\rm M_\odot$ from a Logflat initial mass function \citep{Hirano+2015MNRAS}. The mass of the secondary component ($\rm M_{2\star}$) is determined by assuming a flat mass ratio distribution, where $q=\rm M_{2\star}/\rm M_{1\star}$ is sampled uniformly in the interval $[0.1,\,1)$. \\
Regarding the orbital geometry, the semi-major axis $a$ follows a power-law distribution $\propto a^{-1}$, consistent with Öpik's law, within the range $10 -10^6\,\rm R_\odot$. The orbital eccentricity $e$ is sampled from a standard thermal distribution with probability density $f\,(e)=2\,e$. To ensure dynamical stability and physical consistency of the population, we apply a survival criterion: any system with an initial periapsis separation smaller than 
1.5 times the sum of the stellar radii is excluded from the final database.\\
The physical prescriptions adopted in our models are designed to provide a realistic representation of primordial stellar populations within the \texttt{GAMESH} galaxy formation framework. Although our initial conditions are consistent with recent population studies of Pop III binary populations \citep[e.g.,][]{Tanikawa+2021ApJ, Costa+2023MNRAS, Mestichelli+2024A&A}, the choice of free parameters within BPS codes can significantly affect the final results. Specifically, uncertainties in the efficiency of stellar winds \citep{Torniamenti+2026A&A}, the treatment of the common envelope phase \citep{Giacobbo+2018MNRAS}, and internal mixing -- such as overshooting and convection \citep{Costa+2021MNRAS} -- could broaden the metallicity window for potential GW231123 candidates. This would effectively increase the fraction of star formation available for their production within our simulated over-dense volume.\\
Recent studies exploring the Pop~III channel for GW231123 provide some guidance on how variations in the adopted stellar physics or initial conditions may affect the range of viable progenitors. In particular, \citet{Tanikawa+2025arXiv} and \citet{Popa+2025ApJ} show that binaries with component masses comparable to GW231123 can form from very massive Pop~III stars, although typically at extremely low metallicities
($Z \lesssim 10^{-6}$). Under more optimistic assumptions on internal mixing and nuclear reaction rates, the viable metallicity range may extend up to $Z \sim 10^{-5}-10^{-4}$. This would increase the fraction of star-forming regions capable of producing such systems and therefore the effective SFR
available for their formation within cosmological simulations. \\
Similarly, modifications of the initial orbital architecture may significantly affect the merger delay times. For example,
highly eccentric binaries or systems forming with smaller initial separations can experience much shorter gravitational
wave inspiral times \citep{Peters+1964PhRv}. Several studies find that, under such conditions, delay times can decrease from the
typical values of several Gyr predicted by standard Pop~III binary evolution to values of order $\sim 100\,{\rm Myr}-1\,{\rm Gyr}.$
This would allow binaries formed at the high redshifts predicted by GAMESH to merge within the redshift interval inferred
for GW231123. \\
However, even under these more favorable assumptions, our conclusions remain largely unchanged. Within the \texttt{GAMESH}
framework, the rapid metal enrichment of star-forming halos quickly suppresses the formation of ultra–metal-poor stars.
As a result, the environments capable of producing such Pop~III progenitors contribute only negligibly to the total
SFR in our simulated over-dense volume.

\section{Estimation of the Pop III merger rate upper limit from \texttt{BSEEMP} code}
\label{app:BSEEMP_merger_rate}
In this section we detail the methodology used to derive the upper limit for the merger rate density of Pop III BBHs with masses consistent with GW231123 (GW231123-mass BBHs).
Although GW231123-mass BBHs are produced by both BPS models, those originating from the \texttt{SEVN} code exhibit delay times that exceed the Hubble time, effectively precluding any mergers. To estimate the rate for \texttt{BSEEMP}, we adopted the following procedure.\\
Using the \texttt{BSEEMP} Pop III binary catalog at the lowest metallicity ($Z = 10^{-10}$, i.e., the unique metallicity leading to the formation of GW231123-mass BBHs), we calculated the formation efficiency of GW231123-mass BBHs. This is defined as the number of such systems $(\approx 40)$ per total stellar mass formed:
$$f_{GW} = \frac{N_{GW231123}}{\rm M_{tot, sim}} \approx 1.5 \times 10^{-9} \, \rm M_\odot^{-1}.$$
It is important to emphasize that $f_{GW}$ must be treated as a strict upper limit. In the \texttt{GAMESH} pipeline, binary systems are randomly sampled from BPS catalogs until the total stellar mass of a galaxy is saturated. Given the low formation efficiency of these events, the probability of extracting a GW231123-mass stellar progenitor in a specific galactic environment is $\ll 1\%$.\\
We estimate the upper limit on the birth rate density of these specific progenitors by scaling the Pop III SFR at $Z = 10^{-10}$ with the previously computed formation efficiency:
$$\dot{n}_{form}(t) = f_{GW} \times SFR_{III}(t,\,Z = 10^{-10}).$$
Finally, we combine the upper limit on the birth rate with the average delay time of GW231123-mass BBHs $(\tau_M)$ derived from our BPS catalog in order to obtain the upper limit on the merger rate density of GW231123-mass BBHs:
$$\mathcal{R}(t) = f_{GW} \times SFR_{III}(t - \langle \tau_M \rangle,\,Z = 10^{-10}).$$
The resulting upper limit is shown by the blue curve in Fig. \ref{fig:MergerRateDensity}, where the downward arrow emphasizes its nature as an upper limit. This constraint is approximately eight order of magnitude lower than the merger rate expected from dynamically formed binaries, highlighting the negligible contribution of the isolated Pop III evolution channel to this specific gravitational-wave class of events.

\section{Properties of GW231123-like binary black hole birth environments}
\label{app:birth_places}

\begin{figure*}[h!]
    \centering
    \includegraphics[width=15cm]{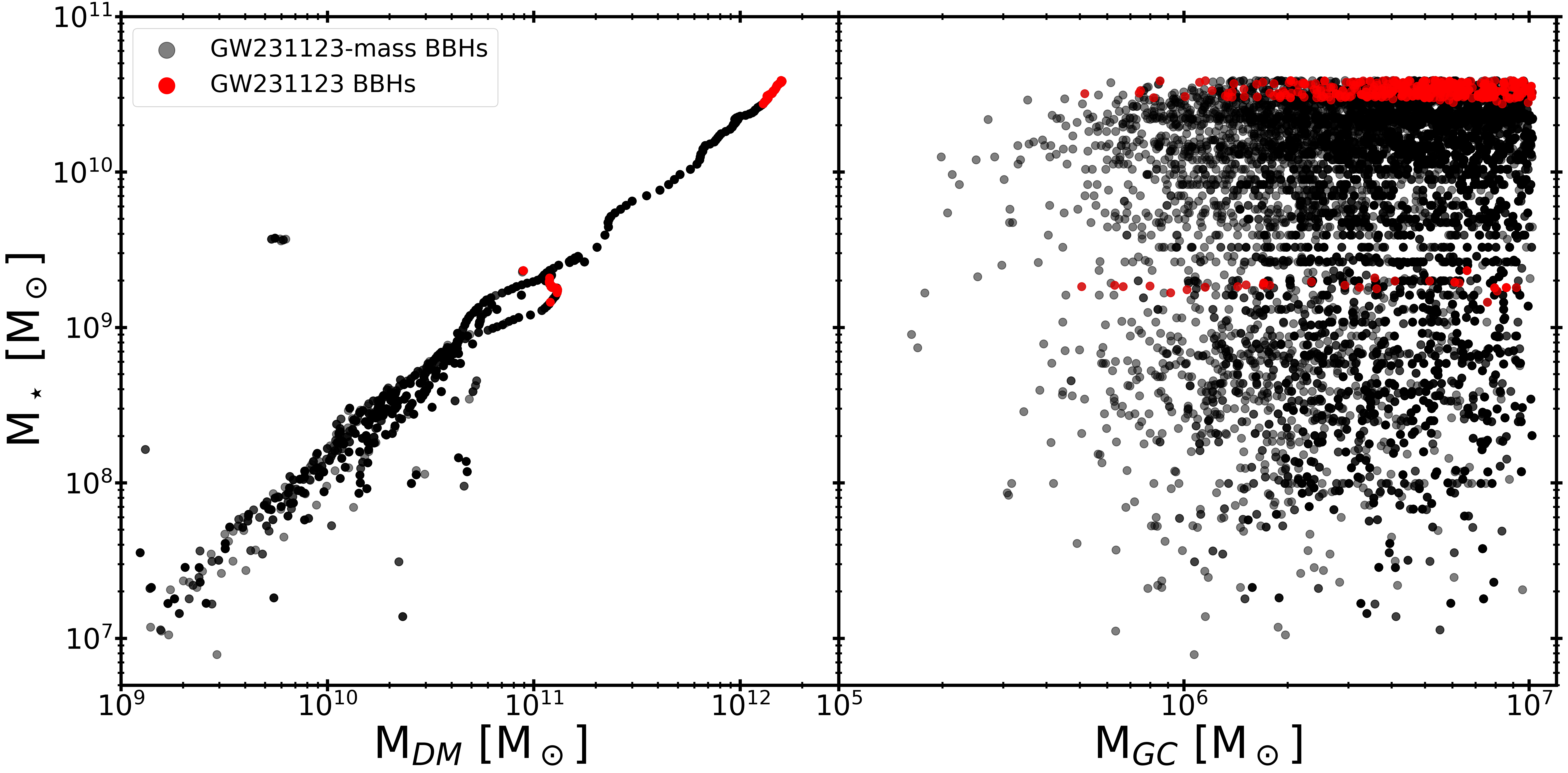}
    \caption{Stellar-halo mass relation for galaxies hosting GW231123-like BBHs (left panel) and  mass distribution of GC progenitors for GW231123-like BBHs (right panel). Black points identify systems hosting GW231123-mass BBHs, selected to match the LVK constraints on the primary and secondary masses, while red points highlight the subset of GW231123 BBHs satisfying the full LVK constraints on mass, spin, and merger redshift.}
    \label{fig:HostGalaxy&Cluster}
\end{figure*}

In Fig. \ref{fig:HostGalaxy&Cluster} we examine the galactic and GC environments hosting GW231123 hierarchical merger candidates. The left panel shows the stellar–halo mass relation of the host galaxies. GW231123-mass BBHs (black points) form across a broad range of galaxy and halo masses ($\rm M_\star\!\sim\!10^7$–$10^{10}\, \rm M_\odot$, ${\rm M_{\rm DM}\! \sim\! 10^9\!-\!10^{12} \, \rm M_\odot}$), whereas the fully consistent GW231123 BBHs (red points) are predominantly hosted by more massive systems ($\rm M_\star \!\!>\!\! 10^9\, \rm M_\odot$, $\rm M_{\rm DM} \!\!>\!\! 10^{11}\,\rm M_\odot$).
This preference reflects their short merger timescales (a few hundred megayears): because formation and coalescence occur in close succession, matching the observed merger redshift of GW231123 (${z = 0.39^{+0.27}_{-0.24}}$) requires formation in relatively massive galaxies at late cosmic times.\\
The right panel shows the relation between host-galaxy stellar mass and parent GC mass ($\rm M_{\rm GC}$). Both GW231123-mass BBHs (black) and the fully consistent GW231123 BBHs (red) preferentially form in massive clusters ($\rm M_{\rm GC}\!\!\gtrsim\!\! 10^6 \, \rm M_\odot$), highlighting the importance of high escape velocities to retain merger remnants against relativistic recoil kicks. However, once the spin and merger-redshift constraints are imposed, the consistent subset shifts toward more massive host galaxies, indicating that the redshift selection primarily drives the preference for higher stellar masses.

\section{Analysis of the spin distribution of GW231123 candidates}
\label{app:spin_analysis}

\begin{figure*}[h!]
    \centering
    \includegraphics[width=15cm]{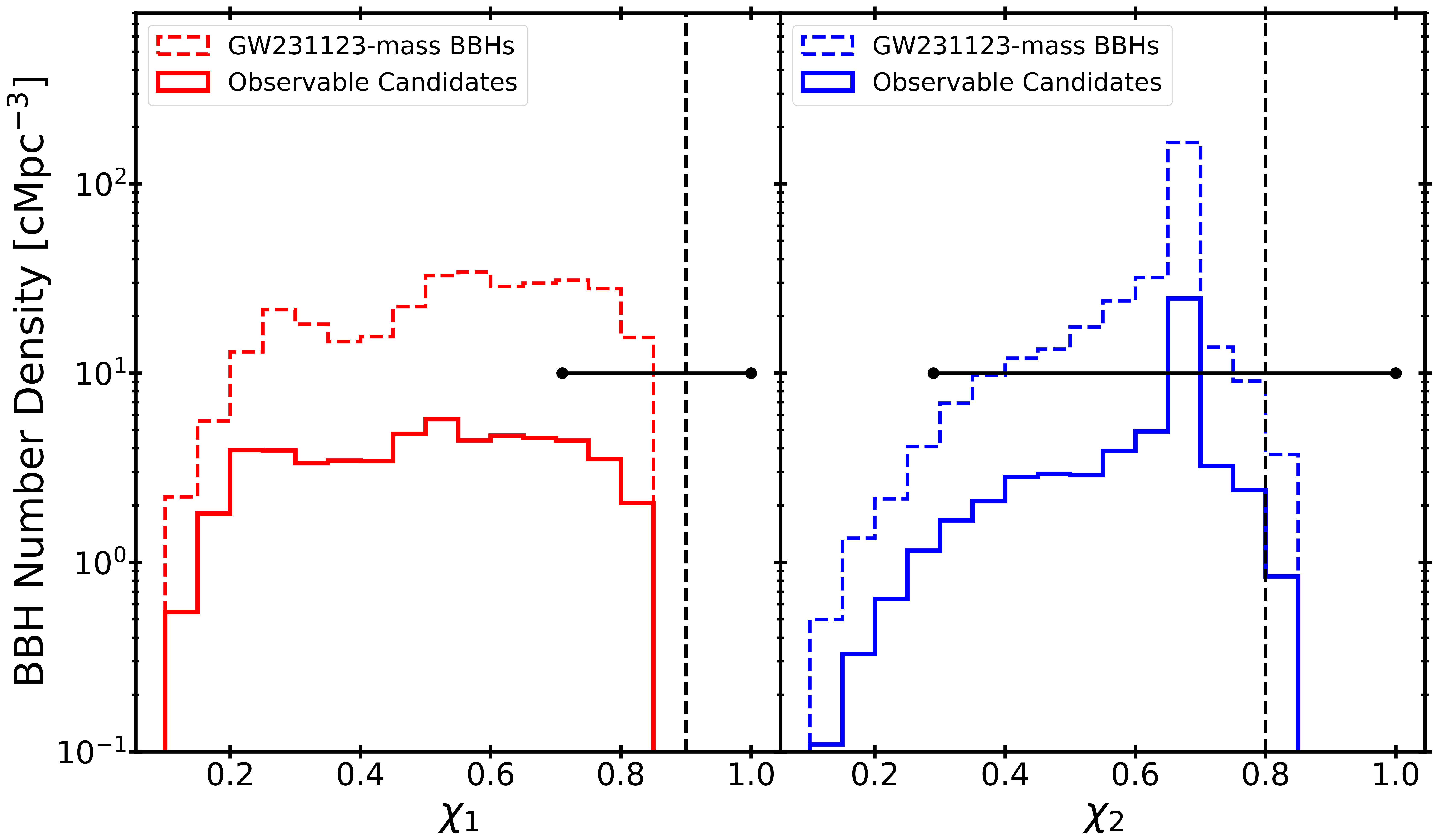}
    \caption{Distribution of the dimensionless spins of the primary ($\chi_1$, left) and secondary ($\chi_2$, right) BHs. Dashed histograms show GW231123-mass BBHs, selected to match the LVK mass priors, while solid histograms represent the subset of these systems which additionally satisfy the merger-redshift constraint. Black vertical lines and horizontal bars mark the LVK median values and 90\% credible intervals for GW231123.}
    \label{fig:BH_Spin}
\end{figure*}

The analysis of dimensionless spins provides further support for the hierarchical origin of GW231123. By comparing the \texttt{RAPSTER} spin distribution with the LVK posteriors, we assess the consistency of our GW231123 candidates.\\
Figure \ref{fig:BH_Spin} shows that the primary spin distribution (left panel) is broad, with many systems spanning $\chi_1 \simeq 0.4-0.8$. The high spin inferred by LVK for GW231123 (black lines and bars) is well reproduced by our hierarchical population, as repeated mergers naturally transfer orbital angular momentum into the remnant spin \citep{Berti+2008ApJ, Gerosa+2021NatAs}. The secondary component (right panel) instead exhibits a pronounced peak around $\chi_2\approx0.7$, also consistent with the observed constraints.\\
As in the mass and metallicity analysis, systems selected to satisfy both the merger-redshift and mass constraints (solid lines) represent a subset of the broader GW231123-mass BBH population (dashed lines). The similarity between the two spin distributions indicate that spin magnitudes show no significant redshift dependence within our hierarchical sample.

\end{appendix}

\end{document}